\documentclass[aps,prl,reprint,showpacs,groupedaddress,amsmath,amssymb,superscriptaddress,nofootinbib]{revtex4-1}% runinaddress, unsortedaddress 
\usepackage{graphicx}% Include figure files
\usepackage[utf8]{inputenc}
\usepackage{dcolumn}% Align table columns on decimal point
\usepackage{url}
\usepackage{subeqnarray}
\usepackage{bm}
\usepackage{csquotes}
\usepackage[hang]{subfigure}
\usepackage[]{lineno}
\usepackage{type1cm}
\usepackage{lettrine}
\usepackage[rawfloats=true]{floatrow}
\usepackage{comment}
\usepackage[colorlinks=true,linkcolor=blue,citecolor=blue,urlcolor=blue, anchorcolor=red, filecolor=red, menucolor=red]{hyperref}
\usepackage{float}
\begin{document}

\title{{Probing Ultrafast Magnetic-Field Generation by Current Filamentation Instability in Femtosecond Relativistic Laser-Matter Interactions}}

\author{G. Raj}
\thanks{G. Raj and O. Kononenko contributed equally to this work.}
\affiliation{LOA, ENSTA Paris, CNRS, Ecole Polytechnique, Institut Polytechnique de Paris, 91762 Palaiseau, France}

\author{O. Kononenko}
\thanks{G. Raj and O. Kononenko contributed equally to this work.}
\affiliation{LOA, ENSTA Paris, CNRS, Ecole Polytechnique, Institut Polytechnique de Paris, 91762 Palaiseau, France}

\author{A. Doche}
\affiliation{LOA, ENSTA Paris, CNRS, Ecole Polytechnique, Institut Polytechnique de Paris, 91762 Palaiseau, France}

\author{X. Davoine}
\affiliation{CEA, DAM, DIF, 91297 Arpajon, France}

\author{C. Caizergues}
\affiliation{LOA, ENSTA Paris, CNRS, Ecole Polytechnique, Institut Polytechnique de Paris, 91762 Palaiseau, France}

\author{Y.-Y. Chang}
\affiliation{Helmholtz-Zentrum Dresden - Rossendorf, Institute of Radiation Physics, Bautzner Landstrasse 400, 01328 Dresden, Germany}

\author{J.~P.~Couperus~Cabadağ}
\affiliation{Helmholtz-Zentrum Dresden - Rossendorf, Institute of Radiation Physics,
Bautzner Landstrasse 400, 01328 Dresden, Germany}

\author{A. Debus}
\affiliation{Helmholtz-Zentrum Dresden - Rossendorf, Institute of Radiation Physics, Bautzner Landstrasse 400, 01328 Dresden, Germany}

\author{H. Ding}
\affiliation{Ludwig-Maximilians-Universit\"at M\"unchen, Am Coulombwall 1, 85748 Garching, Germany}
\affiliation{Max Planck Institut f\"ur Quantenoptik, Hans-Kopfermann-Str. 1, Garching 85748, Germany}

\author{M. F\"orster}
\affiliation{Ludwig-Maximilians-Universit\"at M\"unchen, Am Coulombwall 1, 85748 Garching, Germany}
\affiliation{Max Planck Institut f\"ur Quantenoptik, Hans-Kopfermann-Str. 1, Garching 85748, Germany}

\author{M. F. Gilljohann}
\affiliation{Ludwig-Maximilians-Universit\"at M\"unchen, Am Coulombwall 1, 85748 Garching, Germany}
\affiliation{Max Planck Institut f\"ur Quantenoptik, Hans-Kopfermann-Str. 1, Garching 85748, Germany}

\author{J.-P. Goddet}
\affiliation{LOA, ENSTA Paris, CNRS, Ecole Polytechnique, Institut Polytechnique de Paris, 91762 Palaiseau, France}

\author{T. Heinemann}
\affiliation{Deutsches Elektronen-Synchrotron DESY, 22607 Hamburg, Germany}
\affiliation{Scottish Universities Physics Alliance, Department of Physics, University of Strathclyde, Glasgow G4 0NG, UK}
\affiliation{Cockcroft Institute, Sci-Tech Daresbury, Keckwick Lane, Daresbury, Cheshire WA4 4AD, UK}

\author{T. Kluge}   
\affiliation{Helmholtz-Zentrum Dresden - Rossendorf, Institute of Radiation Physics, Bautzner Landstrasse 400, 01328 Dresden, Germany}

\author{T. Kurz}
\affiliation{Helmholtz-Zentrum Dresden - Rossendorf, Institute of Radiation Physics, Bautzner Landstrasse 400, 01328 Dresden, Germany}
\affiliation{Technische Universit\"at Dresden, 01062 Dresden, Germany}

\author{R. Pausch}
\affiliation{Helmholtz-Zentrum Dresden - Rossendorf, Institute of Radiation Physics, Bautzner Landstrasse 400, 01328 Dresden, Germany}

\author{P. Rousseau}
\affiliation{LOA, ENSTA Paris, CNRS, Ecole Polytechnique, Institut Polytechnique de Paris, 91762 Palaiseau, France}

\author{P. San Miguel Claveria}
\affiliation{LOA, ENSTA Paris, CNRS, Ecole Polytechnique, Institut Polytechnique de Paris, 91762 Palaiseau, France}

\author{S. Sch\"obel}
\affiliation{Helmholtz-Zentrum Dresden - Rossendorf, Institute of Radiation Physics,
Bautzner Landstrasse 400, 01328 Dresden, Germany}
\affiliation{Technische Universit\"at Dresden, 01062 Dresden, Germany}

\author{A. Siciak}
\affiliation{LOA, ENSTA Paris, CNRS, Ecole Polytechnique, Institut Polytechnique de Paris, 91762 Palaiseau, France}

\author{K. Steiniger}
\affiliation{Helmholtz-Zentrum Dresden - Rossendorf, Institute of Radiation Physics, Bautzner Landstrasse 400, 01328 Dresden, Germany}

\author{A. Tafzi}
\affiliation{LOA, ENSTA Paris, CNRS, Ecole Polytechnique, Institut Polytechnique de Paris, 91762 Palaiseau, France}

\author{S. Yu}
\affiliation{LOA, ENSTA Paris, CNRS, Ecole Polytechnique, Institut Polytechnique de Paris, 91762 Palaiseau, France}

\author{B. Hidding}
\affiliation{Scottish Universities Physics Alliance, Department of Physics,
University of Strathclyde, Glasgow G4 0NG, UK}
\affiliation{Cockcroft Institute, Sci-Tech Daresbury, Keckwick Lane, Daresbury, Cheshire WA4 4AD, UK}

\author{A. Martinez de la Ossa}
\affiliation{Deutsches Elektronen-Synchrotron DESY, 22607 Hamburg, Germany}

\author{A. Irman}
\affiliation{Helmholtz-Zentrum Dresden - Rossendorf, Institute of Radiation Physics,
Bautzner Landstrasse 400, 01328 Dresden, Germany}

\author{S. Karsch}
\affiliation{Ludwig-Maximilians-Universit\"at M\"unchen, Am Coulombwall 1, 85748 Garching, Germany}
\affiliation{Max Planck Institut f\"ur Quantenoptik, Hans-Kopfermann-Str. 1, Garching 85748, Germany}

\author{A. D\"opp}
\affiliation{Ludwig-Maximilians-Universit\"at M\"unchen, Am Coulombwall 1, 85748 Garching, Germany}
\affiliation{Max Planck Institut f\"ur Quantenoptik, Hans-Kopfermann-Str. 1, Garching 85748, Germany}

\author{U. Schramm}
\affiliation{Helmholtz-Zentrum Dresden - Rossendorf, Institute of Radiation Physics,
Bautzner Landstrasse 400, 01328 Dresden, Germany}
\affiliation{Technische Universit\"at Dresden, 01062 Dresden, Germany}

\author{L. Gremillet}
\affiliation{CEA, DAM, DIF, 91297 Arpajon, France}

\author{S. Corde}
\email[Corresponding authors: ]{gaurav.raj@polytechnique.edu,\\ olena.kononenko@polytechnique.edu,\\ sebastien.corde@polytechnique.edu.}
\affiliation{LOA, ENSTA Paris, CNRS, Ecole Polytechnique, Institut Polytechnique de Paris, 91762 Palaiseau, France}

%\date{}

\begin{abstract}
We present experimental measurements of the femtosecond time-scale generation of strong magnetic-field fluctuations during the interaction of ultrashort, moderately relativistic laser pulses with solid targets. These fields were probed using low-emittance, highly relativistic electron bunches from a laser wakefield accelerator, and a line-integrated $B$-field of $2.70 \pm 0.39\,\rm kT\,\mu m$ was measured. Three-dimensional, fully relativistic particle-in-cell simulations indicate that such fluctuations originate from a Weibel-type current filamentation instability developing at submicron scales around the irradiated target surface, and that they grow to amplitudes strong enough to broaden the angular distribution of the probe electron bunch a few tens of femtoseconds after the laser pulse maximum. Our results highlight the potential of wakefield-accelerated electron beams for ultrafast probing of relativistic laser-driven phenomena.
\end{abstract}

\maketitle

The Weibel-type current filamentation instability (CFI) \cite{Weibel_PRL_1959, Fried_POF_1959} has been extensively investigated in past decades owing to its recognized importance in an increasing variety of plasma environments. Induced by temperature anisotropies or relative drifts between the plasma constituents \cite{Davidson_POF_1972, Bret_POP_2010,Achterberg_AA_2007a,2019arXiv190409692H}, it gives rise to kinetic-scale, current filaments surrounded by toroidal magnetic fields, through which the charged particles are progressively isotropized \cite{Davidson_POF_1972, Achterberg_AA_2007b, Ruyer_POP_2015a}. This instability is widely thought to underpin the physics of relativistic outflows in powerful astrophysical objects (e.g. gamma-ray bursts, pulsar winds, active galactic nuclei), especially as the source of the collisionless shock waves held responsible for generating nonthermal high-energy particles and radiations \cite{Medvedev_APJ_1999, Spitkovsky_APJ_2008, Kato_APJ_2008, Marcowith_RPP_2016, Lemoine_PRL_2019}. Moreover, it is expected to operate in magnetic reconnection scenarios \cite{Swisdak_APJ_2008}, and has been invoked as a possible generation mechanism for cosmological magnetic fields \cite{Schlickeiser_APJ_2003}.

On the laboratory side, the CFI stands as a key process in intense laser-plasma interactions. In the case of overdense plasmas irradiated at relativistic laser intensities ($I_0\lambda_0^2 \gtrsim 10^{18}\,\mathrm{W\:cm}^{-2}\:\mu\mathrm{m}^2$, where  $I_0$ and $\lambda_0$ are the laser intensity and wavelength, respectively), it arises from the counterstreaming of the forward-directed, laser-accelerated fast electrons and the current-neutralizing, cold plasma electrons \cite{Sentoku_POP_2000, Silva_POP_2002, Ren_POP_2006, Robinson_NF_2014}. The resulting magnetic fluctuations may grow fast enough to cause significant scattering and deceleration of the fast electrons \cite{Sentoku_PRL_2003, Mishra_POP_2008, Adam_PRL_2006, Debayle_PRE_2010}. These effects are generally considered detrimental to fast-electron-based applications, e.g. the fast ignition approach to inertial confinement fusion \cite{Robinson_NF_2014} or target normal sheath ion acceleration \cite{Fuchs_PRL_2003, Gode_PRL_2017, Scott_NJP_2017}. Still, they can also be triggered purposefully in laboratory astrophysics experiments addressing the physics of collisionless shocks, whether involving relativistic laser-solid interactions \cite{Fiuza_PRL_2012, Ruyer_POP_2015b}, laser-driven {interpenetrating plasma flows} \cite{Fox_PRL_2012, Huntington_NP_2015}, or electron beam-plasma interactions \cite{Allen_PRL_2012, Benedetti_NP_2018}.

Experimental evidence for the development of the CFI in relativistic laser-driven plasmas has been mainly provided through characterization of the spatial profiles of the fast electron \cite{Tatarakis_PRL_2003, Wei_PRE_2004, Jung_PRL_2005, Manclossi_PRL_2006} or ion  \cite{Fuchs_PRL_2003,  Metzkes_NJP_2014, Gode_PRL_2017, Scott_NJP_2017, King_HPLSE_2019} beams exiting the target. In situ measurements of the magnetic-field fluctuations at the irradiated target surface have been performed using optical polarimetry \cite{Mondal_PNAS_2012, Chatterjee_NC_2017}, yet this technique cannot access the volumetric distribution of the fields, and the data obtained so far could not capture their femtosecond time-scale dynamics.

\begin{figure}[t]
    \centering
    \resizebox{7.5 cm}{4.5 cm}{\includegraphics[clip=true]{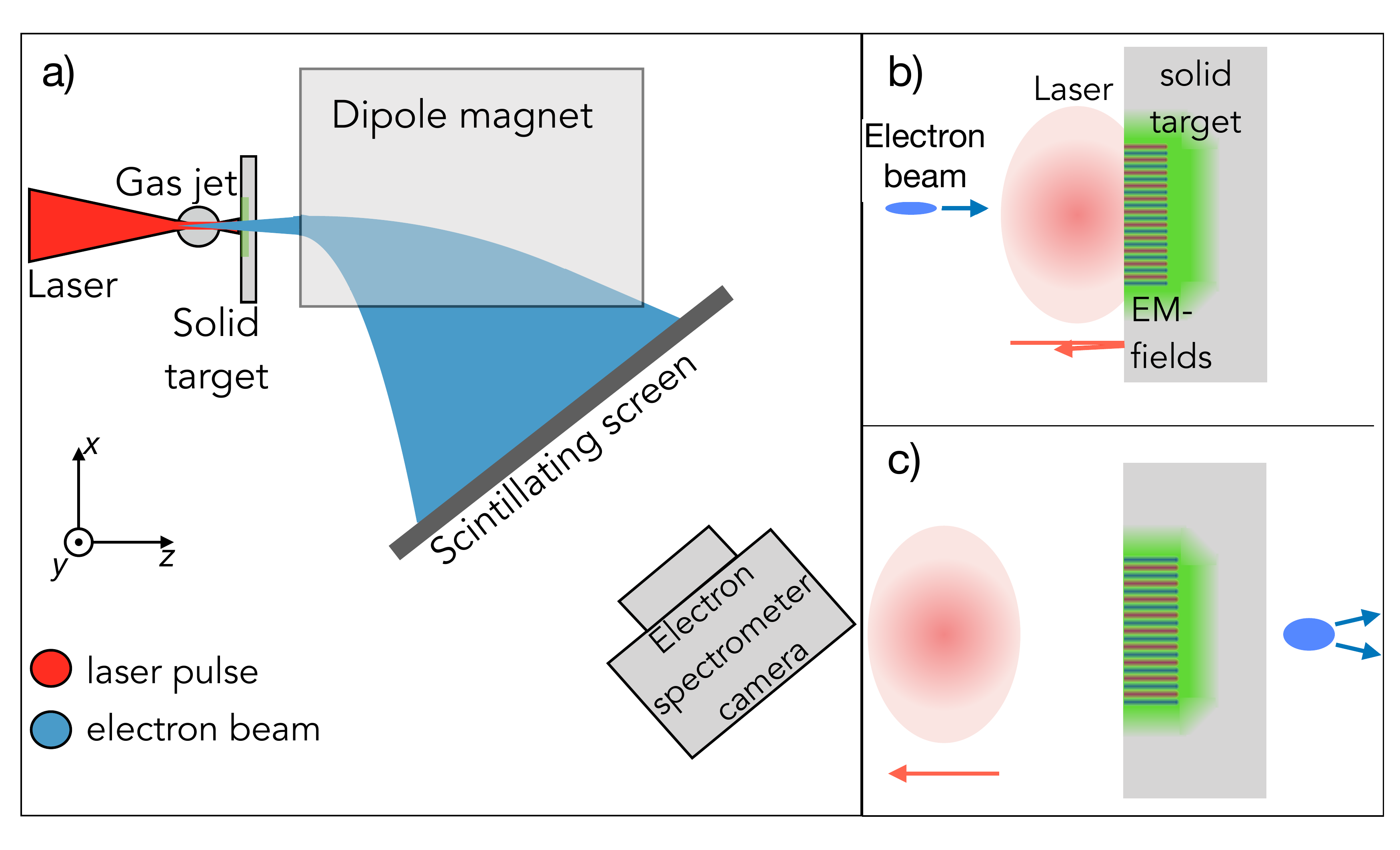}}
    \caption{Schematic of the experimental setup: (a) a laser pulse accelerates a relativistic electron beam from a supersonic gas jet, and is subsequently reflected off a solid foil target placed at the exit of the gas jet. The electron beam passes through the foil and is sent towards an electron spectrometer. (b,c) When traveling across the foil, the beam electrons are scattered by the electromagnetic fluctuations driven by the laser pulse.
    }
    \label{fig1}
\end{figure}

In this Letter, we demonstrate a novel method for diagnosing the kT-level, electromagnetic fluctuations induced in femtosecond laser-solid interactions using an ultrarelativistic probe electron bunch with energies above 100 MeV, produced by a laser wakefield accelerator (LWFA) \cite{Tajima_PRL_1979, Faure_NP_2004, Geddes_NP_2004, Mangles_NP_2004, Goncalves_PRL_2019}. In our experimental setup, the laser pulse driving the LWFA is the same one that induces the electromagnetic fluctuations in a neighboring foil target (see Fig.~\ref{fig1}). This ensures a well-controlled time delay between the electron bunch and the laser pump, and therefore probing of the fluctuations a few tens of femtoseconds only after the on-target laser pulse maximum. Their line-integrated field strength is then inferred from the angular broadening induced upon the electron bunch. Our measurements are supported by 3D fully relativistic particle-in-cell (PIC) simulations, which indicate that the field fluctuations indeed result from a Weibel-type CFI excited at the target surface. Note that ultrafast probing of plasma electromagnetic fields by a LWFA-driven electron beam was previously exploited to image plasma wakefields in LFWAs~\cite{Zhang_PRL_2017} or large-scale, radially expanding magnetic fields in relativistic laser-solid interactions \cite{Schumaker_PRL_2013}.

%%%%%%%%%%%%%%%%%%%%%%%%%%%%%%%%%%%%%%%%%%
%% Beginning of experimental part (Lena)
%%%%%%%%%%%%%%%%%%%%%%%%%%%%%%%%%%%%%%%%%%

The experiment was performed at Laboratoire d’Optique
Appliqu\'ee with the ‘Salle Jaune’ Ti:Sapphire laser system, delivering laser pulses with 30 fs full width at half maximum (FWHM) duration and up to $1.5\,\rm J$ energy on target. The laser pulse had a $810\,\rm nm$ central wavelength and was linearly polarized along the horizontal $x$-axis. Corrected using adaptive optics, it was focused at the entrance of a 3-mm exit diameter gas jet target by a $f/16$ off-axis parabola to a $20\,\rm \mu m$ FWHM spot size in vacuum, yielding a normalized peak vector potential of $a_0\simeq1.5$ when accounting for the experimental intensity distribution in the focal plane. The supersonic gas jet used for the LWFA consisted of a mixture of 99\% hydrogen and 1\% nitrogen, enabling well-controlled electron acceleration through ionization injection \cite{Rowlands_PRL_2008, Pak_PRL_2010, McGuffey_PRL_2010, Doepp_LSA_2017}. Due to relativistic self-focusing and self-steepening in the LWFA stage, the laser field strength is expected to be enhanced to $a_0 \gtrsim 3$~\cite{Corde_NatComm_2013}. After exiting the gas jet, the laser pulse and the electron beam impinged on a thin Mylar or aluminium foil, located at a variable position along the propagation axis. The electron beam transmitted through the foil was characterized by an electron spectrometer comprising a 10-cm-long, 1.0 T dipole magnet deflecting electrons depending on their energy along the horizontal $x$-axis, and a scintillating screen imaged onto a 16-bit camera [see Fig.~\ref{fig1}(a)]. The spectrometer also recorded angular information along the nondispersive vertical $y$-axis (perpendicular to laser polarization), but the large distance (about 35 cm) between the foil and the scintillating screen prevented sub-micron-scale structures of the beam profile close to the target from being resolved.

\begin{figure*}[ht]
    \centering 
    \resizebox{8.55cm}{6.0 cm}{\includegraphics[clip=true]{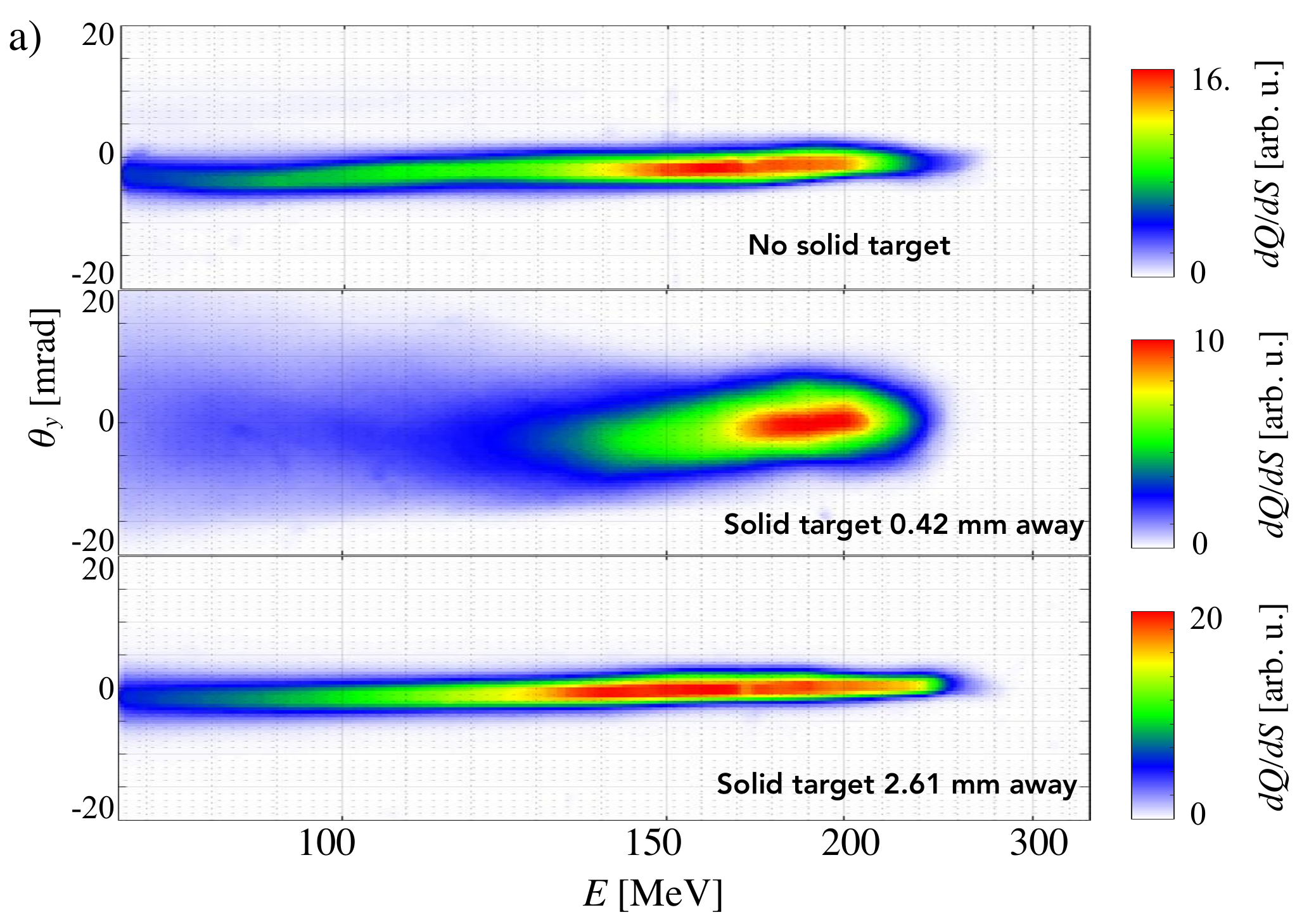}}\hspace{20.0 pt}
    \resizebox{8.55cm}{6.0 cm}{\includegraphics[clip=true]{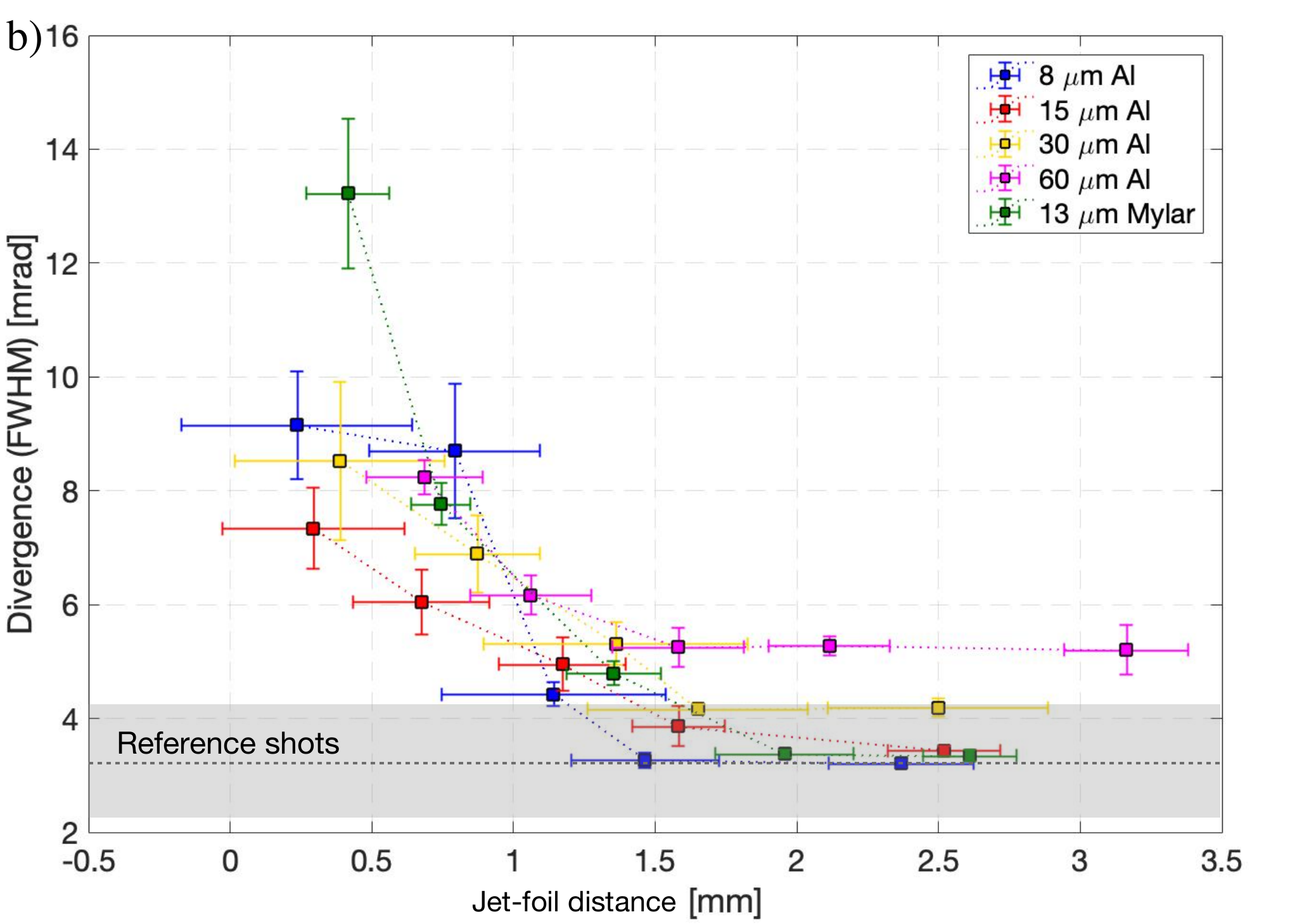}}
    \caption{Experimental results. (a) Typical electron spectra for the reference case (no solid target, top), and for distances of 0.42 mm (middle) and 2.61 mm (bottom) between the 13-$\mu$m-thick Mylar foil and the gas jet exit. (b) Angular divergence (FWHM) of the $150\,\rm MeV$ beam electrons as a function of the distance between the gas jet exit and the solid foil: 8 $\mu$m Al (blue), 15 $\mu$m Al (red), 30 $\mu$m Al (yellow) and 60 $\mu$m Al (magenta), as well as the 13 $\mu$m-thick Mylar foil (green); grey area represents the divergence of the reference shots (no solid target) together with its variation during the experiments.
    }
    \label{fig2}
\end{figure*}

The LWFA was operated in the highly non-linear regime~\cite{Esarey_RMP_2009}, and the electrons from the inner shells of the nitrogen dopant were ionized within the blowout cavity by the high-intensity part of the laser pulse. This resulted in continuous injection as the laser propagated through the gas, and therefore in electron beams with a broad energy spectrum extending beyond 200 MeV [Fig.~\ref{fig2}(a) (top)], a 50-100 pC charge (above 100 MeV) and a 2-4 mrad FWHM divergence. The longitudinal separation between the electron beam and the laser pulse was on the order of the plasma wavelength ($\sim 10\,\mu$m for an electron plasma density of $\sim 10^{19}\,\rm cm^{-3}$). After exiting the gas jet, the peak intensity of the diffracting laser pulse decreased with the propagation distance.

%Seq10Shot86(top): 3.5 mrad; Series average: 3.9 +- 0.5 mrad 
%Seq4Shot4(mid): 10.9 mrad; Series average: 11.7 +- 4.3 mrad 
%Seq10Shot77(bot): 3.5 mrad; Series average: 3.4 +- 0.5 mrad

Figure~\ref{fig2}(a) displays typical electron energy-angle spectra recorded during the experiment. The top panel shows the reference spectrum from the LWFA (no solid target). When a 13-$\rm \mu m$-thick Mylar foil is placed 0.42 mm from the gas jet exit, the beam divergence is significantly increased [Fig.~\ref{fig2}(a), middle]. This effect is strongly reduced when the jet-foil distance is increased to 2.61 mm [Fig.~\ref{fig2}(a), bottom], corresponding to a decrease in the laser intensity on the solid target surface. Multiple scattering of beam electrons in the foil due to elastic collisions cannot account for this behavior since it should cause a negligible increase in the divergence in 13-$\rm \mu m$-thick Mylar (scattering angle of 0.38 mrad for 150 MeV electrons) and be independent of the foil position.

Figure~\ref{fig2}(b) plots the variations of the electron beam divergence with the jet-foil distance (in the range from 0.25 to 3.2 mm) as measured with different targets (13-$\mu$m-thick Mylar and 8 to 60-$\mu$m-thick Al). For each target type, the beam divergence is seen to decrease monotonically with the jet-foil distance. Increasing the Al foil thickness only entails detectable changes at large distances ($\gtrsim 1.5$ mm) due to stronger multiple scattering. These results indicate that the angular broadening of the electron beam takes place in the vicinity of the irradiated target surface.

To assess the possible effect of the foil-reflected laser pulse on the probe electrons, the foil was tilted at $45^\circ$ with respect to the laser axis, so that the reflected laser and the electron beam no longer overlapped. This yielded negligible differences in the observed electron beam divergence, thus proving that the reflected pulse does not account for the increase in the electron beam divergence, and that the source of the latter hardly depends on the laser incidence angle on the target. The only possible scenario, therefore, is that of beam scattering by laser-driven electromagnetic fields within a thin layer behind the target surface.

As a result, our data provide a direct measurement of the integrated Lorentz force experienced by the beam in the solid foil, expressed as an equivalent line-integrated magnetic field,
\begin{align}
\nonumber
B_{x,\mathrm{int}}=\sqrt{\left\langle \left(\int B_x dz\right)^2 \right\rangle_{n_b}},
\end{align}
where the average is weighed by the transverse profile of the electron beam. This field induces a spread $\sigma_{p_y}=eB_{x,\mathrm{int}}$ in the transverse momentum distribution of the electron beam, and therefore contributes to a total divergence $\theta_y^2 = \theta_{y,\mathrm{ref}}^2 + \theta_{y,\mathrm{sc}}^2 + \theta_{y,\mathrm{B}}^2$. Here $\theta_{y,\mathrm{ref}}$ is the original divergence of the LWFA-generated beam, $\theta_{y,\mathrm{sc}}$ is the contribution from the multiple scattering and $\theta_{y,\mathrm{B}}\simeq\sigma_{p_y}/p_z=ecB_{x,\mathrm{int}}/E$ is the contribution from the integrated equivalent magnetic field, with $E$ the electron energy. From the experimentally measured divergence, $\theta_y=13.23\pm1.31$ mrad (FWHM), of the 150 MeV energy electrons passing through the 13-$\mu$m-thick Mylar foil at a 0.42-mm distance, one infers an integrated equivalent magnetic field of $B_{x,\mathrm{int}} = 2.70\pm0.39$ kT$\:\mu\mathrm{m}$.

%%%%%%%%%%%%%%%%%%%%%%%%%%%%%%%%%%%%%%%%%%
%% End of experimental part (Lena)
%%%%%%%%%%%%%%%%%%%%%%%%%%%%%%%%%%%%%%%%%%

%%%%%%%%%%%%%%%%%%%%%%%%%%%%%%%%%%%%%%%%%%
%% Beginning of simulation part (Gaurav)
%%%%%%%%%%%%%%%%%%%%%%%%%%%%%%%%%%%%%%%%%%

\begin{figure*}[ht]
\center\hspace{-20.0 pt}
\resizebox{18.5cm}{10.5 cm}{\includegraphics[clip=true]{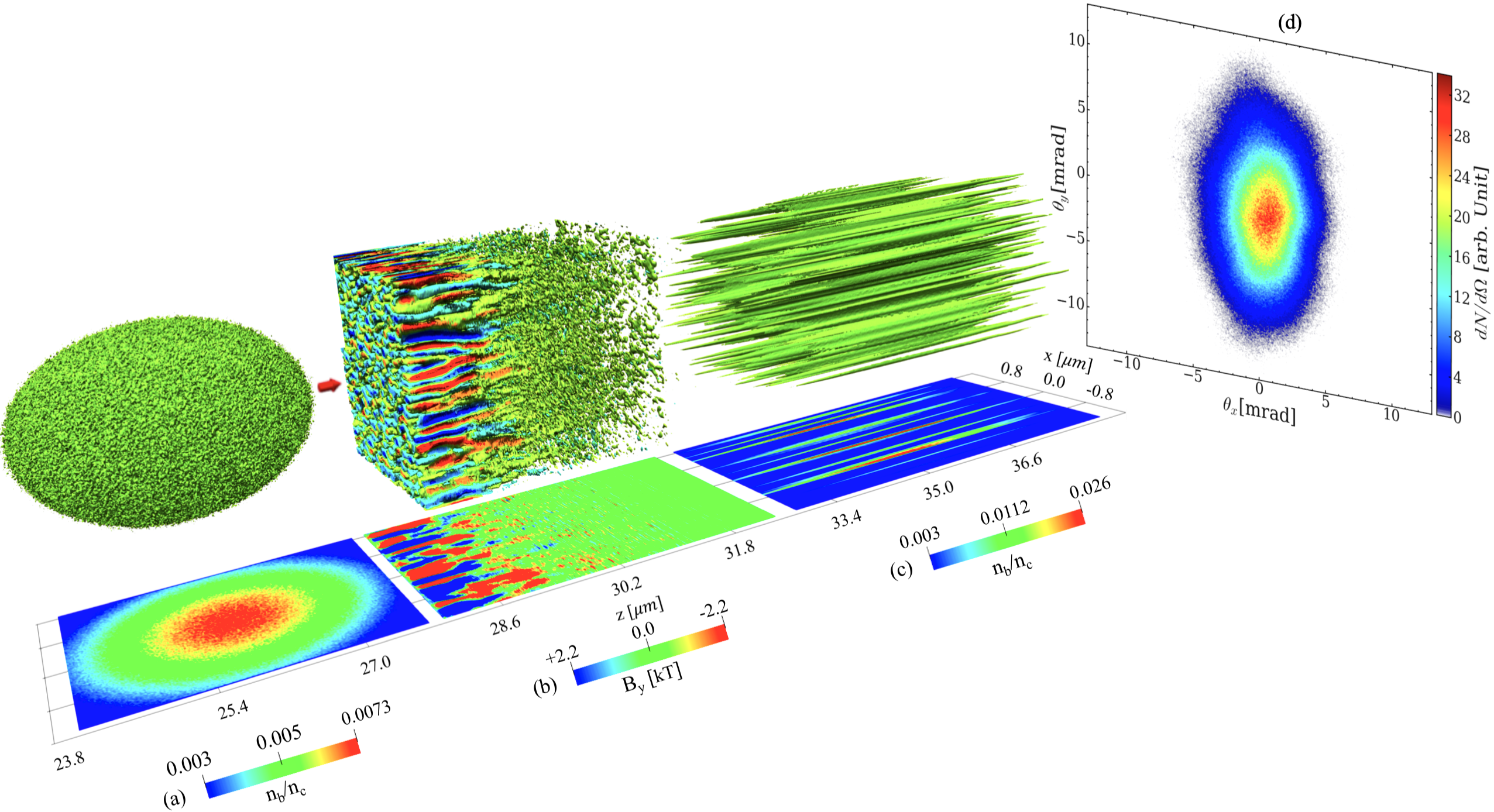}}\vspace{-100 pt}\hspace{0.0 pt}% \vspace{-100 pt}
\raggedleft
\resizebox{5.5 cm}{4.5 cm}{\includegraphics[clip=true]{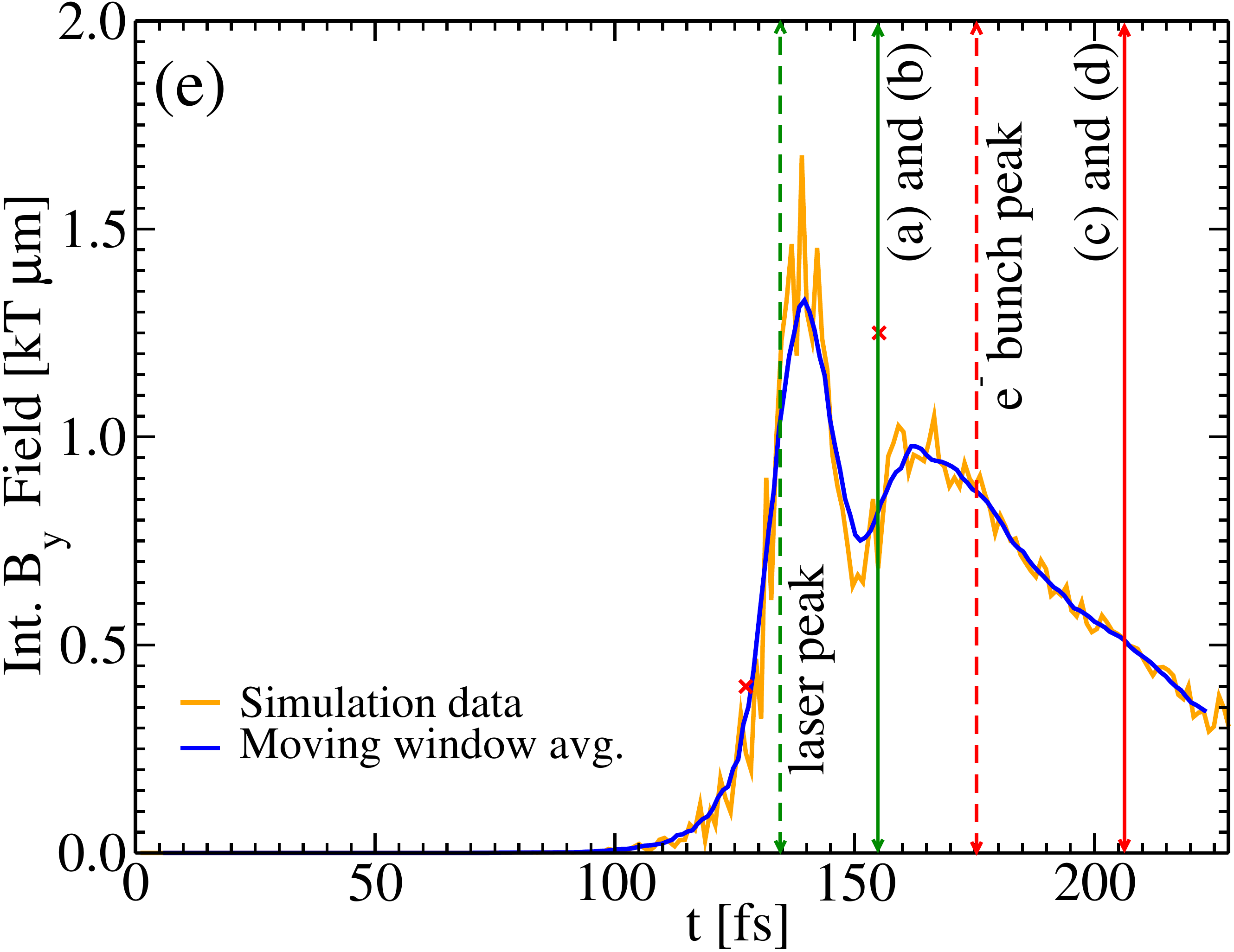}}\hspace{0.0 pt}
\caption{3D PIC simulation snapshots showing isosurfaces and slices at $y=0$ of (a) the electron bunch before entering the Al foil, (b) the $B_y$ component of the magnetic field generated due to laser-solid interaction, (c) the electron bunch after exiting the Al foil. The angular distribution of the final electron bunch is shown in (d), and (e) represents the temporal evolution of the $z$-integrated $B_y$ field obtained from 2D simulations using the same parameters as in the 3D simulation. In (e), the vertical lines indicate the time of arrival at the foil front surface of the peak of the laser pulse (dashed green) and of the electron beam (dashed red), and the time at which (a)-(b) or (c)-(d) snapshots are taken (respectively green and red solid lines). Red crosses in (e) show the instantaneous values of $B_{y,\rm int}$ from the 3D simulation, and the blue curve is a moving window average of the 2D simulation data (orange).
}
\label{fig3}
\end{figure*}

In order to identify the physical mechanism behind the electromagnetic field generation around the target surface, 2D and 3D PIC simulations have been performed using the code \textsc{calder} \cite{Lefebvre_NF_2003, Nuter_POP_2011, Perez_POP_2012, Lobet_JPCS_2016}. These fully relativistic simulations describe both the laser-foil and subsequent beam-plasma interactions, including the effects of binary Coulomb collisions, impact ionization and field ionization. The laser is modeled as a planar wave with a Gaussian temporal profile and a 20 fs FWHM pulse duration. Its field strength on target is estimated to drop from $a_0=2.3$ to 0.7 when the jet-foil distance is increased from 0.5 to 2.7 mm (assuming 1 Joule of laser energy and 15 $\mu$m FWHM spot size at the gas jet exit). The electron beam is initialized with an energy of 150 MeV, a 1 $\mu$m root-mean-square (RMS) transverse size, a 1.6 $\mu$m RMS bunch length, a 50 pC total charge, and a 11 $\mu$m peak-to-peak separation with the laser pulse. The target consists of a 8-$\mu$m-thick, solid-density, neutral plasma of $e^{-}$ and Al$^{3+}$ ions. On its front side is added a linearly ramped preplasma of $0.8\,\rm \mu m$ length to take account of an imperfect laser contrast (see Supplemental Material \cite{Supple19} for a discussion of the weak effect of the preplasma length on the resulting integrated $B$-field). The 3D domain size is $L_x\times L_y \times L_z = 2.1 \times 2.1 \times 45\,\rm \mu m^3$ with a cell size in each direction of $\Delta x = \Delta y =\Delta z = \lambda_0/64$, while for the 2D simulations, the domain size is $L_x \times L_z = 2.1 \times 45\,\rm \mu m^2$ with a cell size $\Delta x =\Delta z = \lambda_0/64$. 50 macro-particles per cell for each species are used in all simulations.

Figure~\ref{fig3} shows results from the 3D PIC simulation for a $0.5\,\rm mm$ jet-foil distance. While several mechanisms may give rise to strong electromagnetic fluctuations in the vicinity of the foil surface (e.g. parametric decay of laser-driven surface oscillations or Rayleigh-Taylor-like instability \cite{Macchi_POP_2002, Kluge_POP_2015}), the Weibel-type CFI appears to be the dominant process under our experimental conditions (see Supplemental Material \cite{Supple19}). The resulting fluctuations, of mainly magnetic nature, exhibit a characteristic filamentary pattern with a $\sim 0.4 \,\rm \mu m$ transverse periodicity, and extending to a $\sim 1 \,\rm \mu m$ depth [Fig.~\ref{fig3}(b)]. The time evolution of the $z$-integrated magnetic field during and after the laser irradiation is presented in Fig.~\ref{fig3}(e), showing that the beam electrons experience fully-grown magnetic fields as soon as they enter the target. Their $(\theta_x, \theta_y)$ angular distribution after transiting through the target is displayed in Fig.~\ref{fig3}(d): the beam divergence along the vertical ($y$) direction is measured to be $\theta_y \simeq 10\,\rm mrad$ (FWHM), much larger than its initial value ($\simeq 0.1\,\rm mrad$) in the simulation. Moreover, these magnetic deflections translate into strong transverse modulations in the beam profile [compare Fig.~\ref{fig3}(a) and Fig.~\ref{fig3}(c)]. The asymmetry between the horizontal ($x$) and vertical ($y$) divergences originates from the stronger laser-induced electron heating along the laser polarization axis ($x$); this excites current modulations preferentially along the cold ($y$) axis, hence leading to $B_{x,\rm int} > B_{y,\rm int}$ and to a larger vertical divergence. 
%Also, the simulation confirms that the reflected laser pulse has no observable effect on the electron beam divergence.

\begin{figure}[t]
\centering\hspace{0 pt}
\resizebox{8.5cm}{5.6cm}{\includegraphics[clip=true]{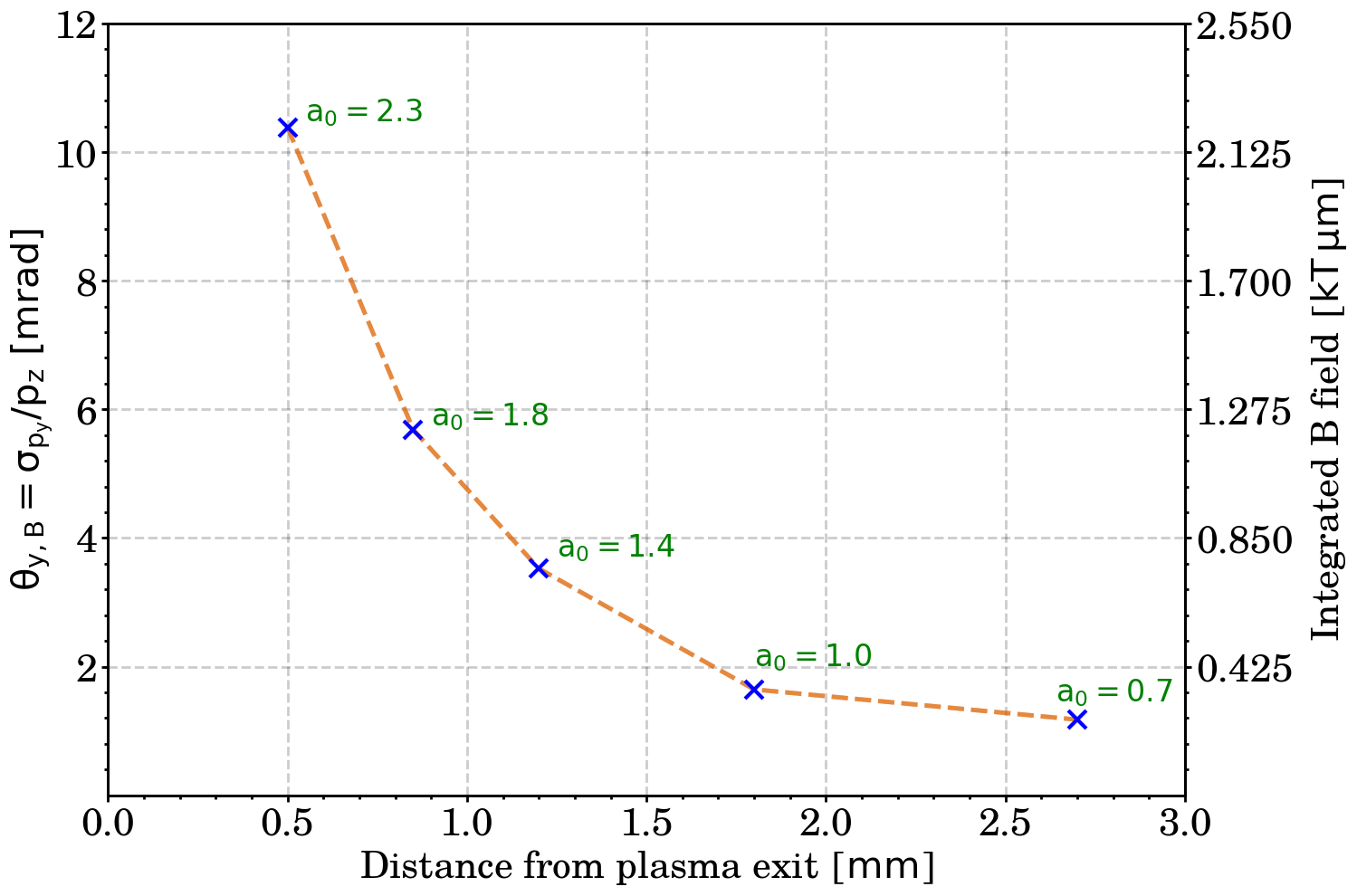}}%
\caption{Parameter scan using 3D PIC simulations, showing the integrated magnetic field, $B_{x,\rm int}$, as experienced by $150\,\rm MeV$ electrons passing through a 8-$\rm \mu m$-thick Al foil placed at variable distance from the gas jet (in green is indicated the corresponding laser field strength), and the resulting contribution $\theta_{y,\rm B}$ to the vertical divergence (FWHM).}
\label{fig4}
\end{figure}

To further compare the simulation results with the experimental observations, we plot in Fig.~\ref{fig4} the results of a parametric scan using 3D PIC simulations where the jet-foil distance is varied from 0.5 to $2.7\,\rm mm$, corresponding to an estimated $a_0$ ranging from 2.3 to 0.7. The integrated magnetic field experienced by the electron beam and the resulting angular broadening are observed to monotonically decrease as the foil is moved away from the gas jet exit, in good agreement with the experimental measurements [Fig.~\ref{fig2}]. This shows the sensitivity of the CFI-induced magnetic fluctuations to the intensity of the femtosecond laser drive pulse.

%%%%%%%%%%%%%%%%%%%%%%%%%%%%%%%%%%%%%%%%%%
%% End of simulation part (Gaurav)
%%%%%%%%%%%%%%%%%%%%%%%%%%%%%%%%%%%%%%%%%%

To conclude, we have demonstrated, both experimentally and numerically, the potential of low-emittance, LWFA-generated electron beams to probe the submicron-scale magnetic fields induced by a {Weibel-like} current filamentation instability developing during ultraintense laser-solid interactions. Supported by 3D PIC simulations, our measurements have allowed us to infer the generation of an integrated magnetic field strength $B_{x,\rm int} = 2.70 \pm 0.39\,\rm kT \,\mu  m$ around the front surface of a solid target irradiated by a $\sim$ 20 fs, $\sim 10^{19}$ W$\:$cm$^{-2}$ laser pulse. These results pave the way for time-resolved $B$-field measurements at femtosecond time scales, by generating the probe electron beam from an auxiliary laser pulse with controlled delay. Our results are also of prime interest for staged plasma-based accelerators \cite{Steinke_Nature_2016}, including novel hybrid schemes, which aim to miniaturize beam-driven plasma wakefield accelerators (PWFA) \cite{Chen_PRL_1985,Litos_Nature_2014} and achieve unprecedented beam quality by using relativistic electron drive beams from a LWFA \cite{Hidding_PRL_2010, Chou_PRL_2016, Ferri_PRL_2018, Gilljohann_PRX_2019, DelaOssa_PTRS_2019}, and separating the LWFA and PWFA by a thin foil. The present study highlights the need to mitigate the CFI (e.g. by depleting the laser pulse energy before it hits the solid target) so as to avoid degrading the quality of the electron beam driving the subsequent acceleration stage.

\begin{acknowledgments}
This work was supported by the European Research Council (ERC) under the European Union’s Horizon 2020 research and innovation programme (Miniature beam-driven Plasma Accelerators project, Grant Agreement No. 715807). H. D., A. D., M. F., M. F. G., and S. K. were supported by DFG through the Cluster of Excellence Munich Centre for Advanced Photonics (MAP EXC 158). Numerical simulations were performed using HPC resources from PRACE (Grant No: 2017174175) and GENCI-TGCC (Grants 2018-A0040507594 and 2019-A0060510786) with the IRENE supercomputer.
\end{acknowledgments}

%\bibliography{Bibsh}


\end{document}